\documentclass[aip,apl,reprint]{revtex4-1}
\usepackage{graphicx}
\usepackage{dcolumn}
\usepackage{bm}

\begin{document}
\title{Spontaneous emission control of single quantum dots by electromechanical tuning of a photonic crystal cavity} %Title of paper
\author{L. Midolo}
\email[]{l.midolo@tue.nl}
\author{F. Pagliano}
\author{T. B. Hoang}
\author{T. Xia}
\author{F. W. M. van Otten}
\affiliation{COBRA Research Institute, Eindhoven University of Technology, P.O. Box 513, NL-5600MB Eindhoven, The Netherlands}
\author{L. H. Li}
\author{E. Linfield}
\affiliation{School of Electronic and Electrical Engineering, University of Leeds, Leeds LS2 9JT, United Kingdom}
\author{M. Lermer}
\author{S. H\"{o}fling}
\affiliation{Technische Physik and Wilhelm Conrad R\"{o}ntgen Research Center for Complex Material Systems, Universit\"{a}t W\"{u}rzburg, Am Hubland, D-97074 W\"{u}rzburg, Germany}
\author{A. Fiore}
\affiliation{COBRA Research Institute, Eindhoven University of Technology, P.O. Box 513, NL-5600MB Eindhoven, The Netherlands}

\date{\today}

\begin{abstract}
We demonstrate the control of the spontaneous emission rate of single InAs quantum dots embedded in a double-membrane photonic crystal cavity by the electromechanical tuning of the cavity resonance. Controlling the separation between the two membranes with an electrostatic field we obtain the real-time spectral alignment of the cavity mode to the excitonic line and we observe an enhancement of the spontaneous emission rate at resonance. The cavity has been tuned over 13 nm without shifting the exciton energies. A spontaneous emission enhancement of $\approx 4.5$ has been achieved with a coupling efficiency of the dot to the mode $\beta\approx92\%$.
\end{abstract}

\maketitle 

\begin{figure}
\includegraphics{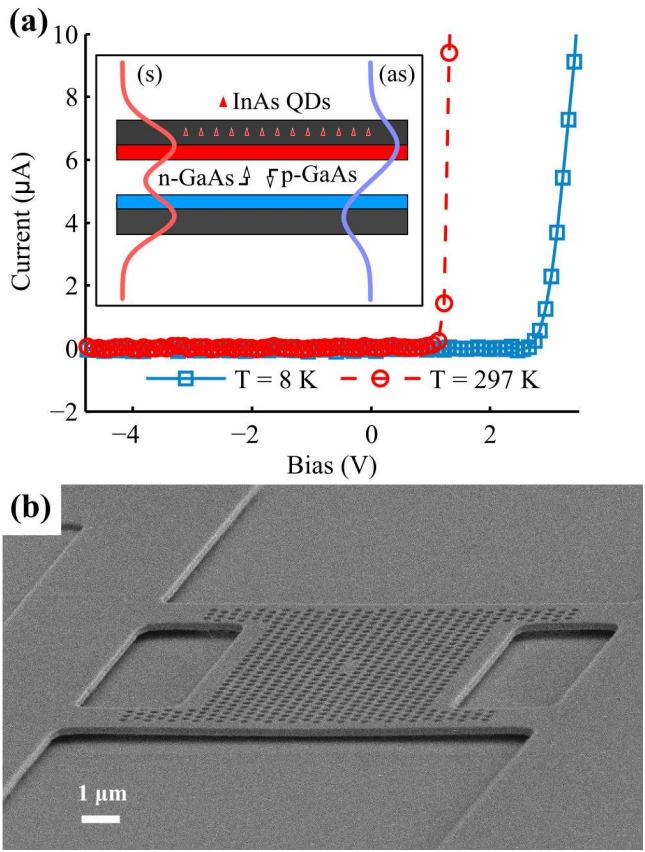}%
\caption{\label{fig:sem}(a) IV curve of the p-i-n diode at room (red circles) and low (blue squares) temperatures. Inset: sketch of the double-membrane showing the doped layers, the symmetric (s) and the anti-symmetric (as) mode profiles. (b) Scanning electron microscope (SEM) image of an L3 cavity realized on a double-membrane GaAs structure (contacts are not visible). The structure is a $12\times12 \mu\text{m}^2$ bridge with four 2-$\mu$m-long suspension arms to increase the flexibility. The photonic crystal has a lattice constant $a=370$ nm and a hole radius $r=0.31 a$. }%
\end{figure}

The coupling of a quantum emitter such as a quantum dot (QD) to a semiconductor photonic crystal cavity (PCC) has shown to be a promising method to realize single photon sources on a chip,\cite{shields_semiconductor_2007} enabling applications in quantum key distribution and quantum photonic integrated circuits (QPIC). 
Two-dimensional PCCs are commonly used for this purpose due to the high achievable Q factors and small mode volumes.\cite{akahane_high-q_2003,song_ultra-high-q_2005} 
The spontaneous emission rate of a two-level system is strongly affected by the local density of optical states provided by the surrounding electromagnetic resonator\cite{purcell_purcell_1946} and can be enhanced or suppressed depending on the spectral alignment between emitter and cavity. 
The spectral control of QDs has been already achieved using different methods such as temperature tuning,\cite{gevaux_enhancement_2006,faraon_local_2007} Stark effect\cite{hogele_voltage-controlled_2004} and strain tuning,\cite{seidl_effect_2006} while the control of the cavity resonance is more challenging. Cavity tuning has been obtained by controlled gas adsorption and local heating,\cite{mosor_scanning_2005,rastelli_situ_2007} however this technique produces a permanent change in the QD emission energy preventing the separate control of QD and cavity. For QPIC applications, where many devices have to operate at the same wavelength, it is essential to tune each cavity independently over a wide wavelength range ($>10$ nm), without affecting the QD emission wavelength and the Q factor. 
An attractive solution which fulfills all these requirements is provided by nano-opto-electro-mechanical structures (NOEMS). 
Previous works have demonstrated reconfigurable PCCs based on the electrostatic actuation of laterally coupled nanobeams,\cite{perahia_electrostatically_2010,frank_programmable_2010} slotted cavities \cite{winger_chip-scale_2011} and two-dimensional PCCs on double membranes.\cite{notomi_optomechanical_2006,midolo_electromechanical_2011} The use of a vertically-coupled double-membrane is particularly convenient since it allows us to separate the QD layer from the actuation region in the vertical direction, removing any possible interaction between the electrostatic field and the QDs. When two PCCs are brought at small distances (see inset Figure \ref{fig:sem}(a)) they couple evanescently, producing a splitting in a symmetric and an antisymmetric mode which shift in wavelength depending on the distance between the membranes. This technique has been demonstrated on InGaAsP/InP,\cite{midolo_electromechanical_2011} and has been shown not to affect the cavity Q.\cite{notomi_optomechanical_2006} However the operation at cryogenic temperatures (which is fundamental for QPIC applications) and the tuning to a single quantum dot have not yet been shown.
In this paper we demonstrate the electromechanical control of the spectral alignment of a cavity mode to single QDs at low temperatures, using a double-membrane structure. 

The device is fabricated on a sample grown by molecular beam epitaxy on an undoped (100) GaAs substrate. It consists of two GaAs layers (thickness 160 nm) and a 200-nm-thick Al$_{0.7}$Ga$_{0.3}$As sacrificial layer in between. A $1 \mu$m thick Al$_{0.7}$Ga$_{0.3}$As sacrificial layer isolates the double-membrane structure from the substrate. 
The upper 50-nm-thick region of the lower membrane is p-doped whereas the lower 50-nm-thick region of the upper membrane is n-doped ($n=p=3\cdot10^{18}$ cm$^{-3}$). In the upper membrane, above the n-doped region, a single layer of low-density self-assembled InAs quantum dots, emitting at 1300 nm at 300 K is grown.\cite{alloing_growth_2005} Fabrication starts by defining contact vias to both membranes by optical lithography and wet etching of GaAs in citric acid/peroxide (C$_6$H$_8$O$_7$ monohydrate mixed 1:1 with water by weight) and Al$_{0.7}$Ga$_{0.3}$As in HF 1\%. A beam-shaped structure (shown in Fig. \ref{fig:sem}(b)) is etched, together with alignment marks, down to the p-layer, situated on the lower membrane. Ti/Au (50/200 nm) pads are evaporated to form metal contacts to both doped layers. 
The photonic crystal design is patterned by 30kV electron beam lithography, transferred to a  400-nm-thick Si$_3$N$_4$ mask layer and deeply etched into both membranes by inductively coupled plasma (Cl$_2$:N$_2$ chemistry at 200 $^\circ$C). Before removing it, the residual nitride mask is used as structural support to stiffen the bridge and avoid stiction phenomena through the undercut process.\cite{midolo_electromechanical_2011} The inter-membrane and the underlying Al$_{0.7}$Ga$_{0.3}$As sacrificial layers are first de-oxidized in HF 1\%, then etched in hydrocloric acid (36\% concentration) at 2$^\circ$C and finally cleaned in HF 5\%. 
Finally the sample is rinsed in hot isopropanol (to reduce the liquid surface tension) and the nitride mask is removed by CF$_4$ plasma. The final device is shown in Fig. \ref{fig:sem}(b).
The room temperature current-voltage (IV) curve of the device in Fig. \ref{fig:sem}(a) (red circles) shows a good p-i-n junction behavior with a reverse current as low as -6 nA. The high turn on voltage is due to the Schottky nature of the Ti/Au contacts on the n-layer which, however, does not affect the operation of the actuator. 

\begin{figure}
\includegraphics{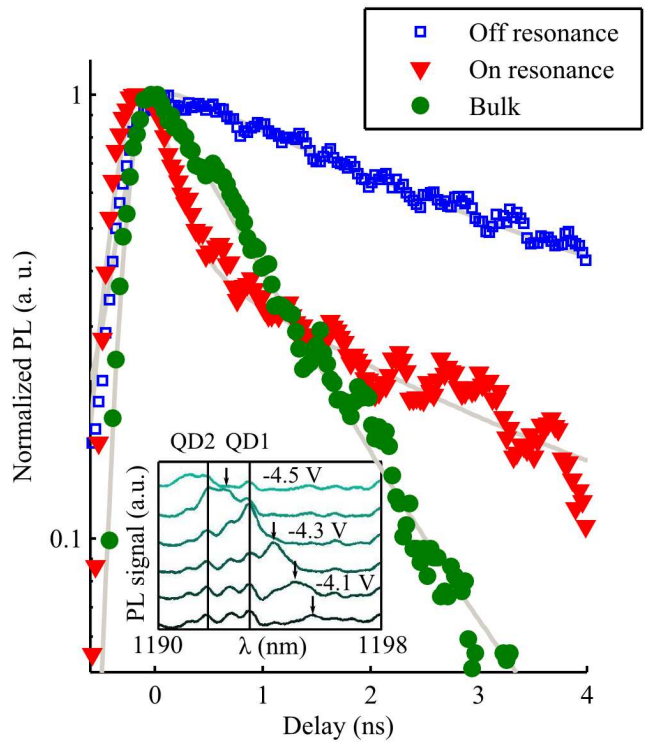}%
\caption{\label{fig:tr} Time-resolved PL of the QD located at 1193.3 nm when the mode is on (red triangles) and off (blue squares) resonance. The decay curve of the QD ensemble in the bulk is also shown as a comparison (green circles). The on-resonance PL is measured with a pump power $P=20$ nW whereas the power $P=90$ nW is used for the off-resonance and the bulk emission measurement. The offset has been subtracted from the data and the curves have been normalized for clarity. The inset shows the PL spectra at $P=20$ nW as the mode (indicated by the arrows) is crossing the excitonic line(s).}%
\end{figure}

The sample is placed in a He-flow cryostat equipped with electrical probes and cooled down to 8 K (as measured on the sample holder). At this temperature the IV curve of the diode (\ref{fig:sem}(a) blue squares) shows an increased turn-on voltage (from 1.2 V to 2.7) and a similar behavior in reverse bias as compared to room temperature.
While cooling and warming up the device it is crucial to avoid condensation of air molecules on the sample surface. Even if the chamber is kept in vacuum a small amount of condensed water can form liquid bridges between the membranes which will eventually lead to stiction failures, especially during the warming process. Using heating pads located near the sample, condensation is avoided and thermal cycling is possible without harming the devices.
The PL spectra are acquired at low temperature using a 80 MHz pulsed diode laser ($\lambda = 757$ nm, pulse width $\sim 70$ ps), focused on the sample with an objective (numerical aperture 0.4).
\begin{figure}
\includegraphics{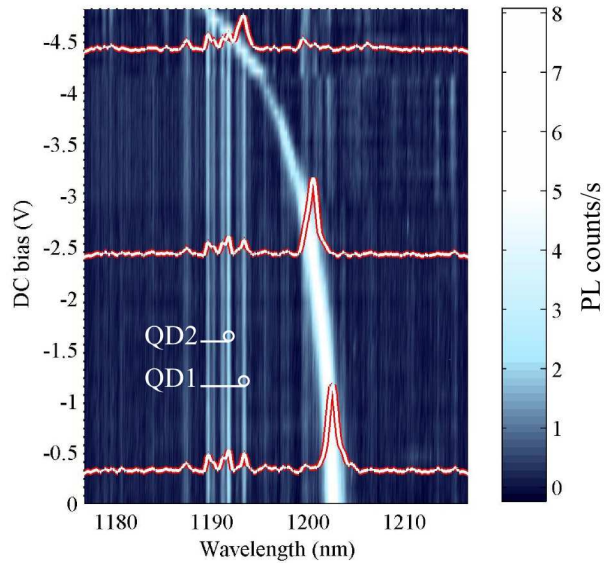}%
\caption{\label{fig:pl} Low-temperature (8 K) PL of an anti-symmetric L3 mode as a function of the applied DC bias using an average pump power $P=40$ nW. The L3 mode is gradually shifted in resonance to the QD lines. The spectrum is obtained sweeping the bias from the 0 to -3 V first, then from -3 to -4.2 V and finally from -4.2 to -4.8 V. Every time the bias is set to 0 V and the original cavity wavelength $\lambda_0$ is restored.}%
\end{figure}
The cavity discussed in the following was made by removing three holes (L3) from a triangular lattice of holes with period $a=370$ nm and fill factor $\text{FF}=0.35$ as measured with SEM analysis.  
In Figure \ref{fig:pl}, the PL of the cavity when a low voltage is applied and with low excitation power (bottom white line) is shown. The cavity mode, located at $\lambda_0=1202.6$ nm, is visible even if it is not on resonance with an emitter, since it is coupled to the spectrally broad emission originated by the QDs.\cite{winger_explanation_2009,chauvin_controlling_2009} Since the QD ground state emission shifts approximately 100 nm to shorter wavelengths when the temperature is lowered, the cavity has been designed to have the $y$-polarized anti-symmetric mode resonant around 1200 nm. The Q factor is $\approx 1100$ which is in good agreement with the theoretical value obtained by finite element (FE) simulations $(\text{Q}_{as}=1300)$. Several excitonic lines, clearly off-resonance, are visible around 1190 nm. These dots are located in the cavity region as it is verified by moving the PL pump laser. As the reverse bias across the membranes is increased, the cavity blue-shifts over a 13 nm range with a maximum applied bias of -4.8 V. The voltage is not increased further to avoid pull-in. 
The tuning is fully reversible and reproducible. In fact, the spectra shown in Fig. \ref{fig:pl} have been collected using several bias ranges to compensate for setup drifts and no discontinuity or hysteresis is observed in the mode tuning. Moreover when a bias is set, the spectral position of the mode is stable within few tenths of nm over a long time ($> 30$ minutes). No effect from the tuning is visible on the measured Q factor, as expected.\cite{notomi_optomechanical_2006} Stark-induced tuning of excitonic lines is not observed within the QD linewidth (approximately $200 \mu$eV with the spectrometer resolution used here) which indicates that any possible residual electric field applied on the QDs is negligible. 
When the mode reaches the excitonic lines, a modulation of intensity is observed, indicating a coupling between the dots and the mode. This is evidenced more clearly at the lower pumping power of $\approx 20$ nW, (see inset of Fig. \ref{fig:tr}), where the mode emission is further suppressed and an enhancement of the excitonic lines is observed as the mode crosses them.
To further confirm the QD-cavity coupling, a time-resolved PL experiment is performed on QD1 ($\lambda = 1193.3$ nm) and QD2 ($\lambda = 1191.6$ nm) as indicated in Fig. \ref{fig:pl}. The PL signal is dispersed into a $f = 1$ m spectrometer, filtered with an exit slit (bandwidth $\approx 0.5$ nm) and sent to a superconducting single photon detector (SSPD).\cite{goltsman_picosecond_2001} The instrument response function (IRF), measured from the laser, indicates a system temporal resolution of 190 ps. All the decay histograms have been fitted with the sum of one or two exponentials. To take into account the response of the system, the fit model is obtained from the convolution of the (bi-)exponential decay with the measured IRF.
Figure \ref{fig:tr} shows the results of the time-resolved experiment for QD1 and the corresponding fits.
The emission from the ensemble of QDs in the bulk (away from the cavity) is measured first (filled green circles), to obtain the reference lifetime. A single exponential decay with a lifetime $\tau_{\text{bulk}} = (1.08\pm0.05)$ ns is obtained which is in good agreement with the values previously obtained on similar QDs.\cite{zinoni_time-resolved_2006,balet_enhanced_2007} 
Then the decay rate of QD1 off-resonance (no dc bias) is measured (squares) resulting in single exponential decay with $\tau^{\text{QD1}}_{\text{off}} = (3.3\pm0.1)$ ns. The slow decay rate is an indication of the suppression of the emission rate due to the absence of available optical states in the photonic crystal. The voltage is gradually increased to -4.3V, where the mode is resonant with QD1, and the laser power is reduced to suppress the emission of the mode. The decay of the dot on resonance clearly shows a bi-exponential behavior. The fast decay component ($\tau^{\text{QD1}}_{\text{on}} = (0.24\pm0.05)$ ns) indicates the rate enhancement of the exciton due to Purcell effect while the slow decay ($\tau_{\text{BG}} = (2.5\pm0.1)$ ns) is attributed to the spin flip transition from dark to bright exciton.\cite{johansen_probing_2010}
The measured spontaneous emission enhancement is $\tau_{\text{bulk}}/\tau^{\text{QD1}}_{\text{on}} = (4.5\pm 1)$. The high uncertainty comes from the error on the fitted lifetime on resonance. By correcting for the emission rate into leaky modes we derive the emission rate into the mode $(1/\tau^{\text{QD1}}_{\text{on}}-1/\tau^{\text{QD1}}_{\text{off}})$ which correspond to a Purcell factor $F_p = \tau_{\text{bulk}}\times(1/\tau^{\text{QD1}}_{\text{on}}-1/\tau^{\text{QD1}}_{\text{off}}) = (4.2 \pm 1)$. 
From the value of Q, an indication of the maximum achievable Purcell factor, assuming an emitter with the best spatial alignment, can be extracted with the formula $F_p = (3/4\pi^2) (\lambda/n)^3 Q/V_{\text{eff}} = 73$, where $V_{\text{eff}}=1.14 \times (\lambda/n)^3$ is the effective mode volume obtained from FE simulations of the anti-symmetric mode of the L3 cavity and $n=3.42$ is the refractive index of GaAs. 
The observed $F_p$ value is much lower than the theoretical one probably because of the limited spatial alignment. 
The second QD can also be easily addressed by increasing the voltage to -4.5 V and it shows $\tau^{\text{QD1}}_{\text{on}} = (0.52\pm0.07)$ ns and a rate enhancement of $2.1\pm0.5$. 
We note that for this pump power level, the lowest for which time-resolved PL could be taken, the mode pumping by the background contributes by only about 15\% to the intensity measured from the QD1 line on resonance. Additionally, the fact that the measured $\tau^{\text{QD1}}_{\text{on}}$ is much shorter than the measured decay time of the mode line off resonance ($\tau_{\text{mode}} = 0.46 \pm 0.02$ ns) confirms that the time-resolved PL signal originates primarly from the dot and not from the background feeding the mode.
The coupling efficiency of the dot to the cavity mode is given by the $\beta$-factor: $\beta = 1 - (\tau^{\text{QD}}_{\text{on}}/\tau^{\text{QD1}}_{\text{off}})$. For QD1 $\beta \approx 92\%$ and for QD2 $\beta \approx 87\%$. These values are comparable to what has been obtained before on L3 cavities \cite{balet_enhanced_2007} and photonic crystal waveguides.\cite{ba_hoang_enhanced_2012} It indicates that the double-membrane NOEMS can be used as the basis for the realization of efficient single photon sources where the control of the spontaneous emission is obtained in real-time, without affecting the properties of the surrounding electromagnetic mode or the energy of the emitter.

To summarize, we have presented a method to controllably tune the mode of a PCC around the emission energy of single QDs at low temperature. A tuning range of 13 nm has been obtained at 8 K, allowing the reproducible coupling to different excitonic lines. Additionally, a four-fold enhancement in the emission rate of the exciton has been measured, together with a coupling efficiency $\beta\approx92\%$. The spectral tuning of the cavity is obtained without altering the energy of the QD excitons which is a key result for applications in solid state single photon sources and for CQED experiments. Future work will address the possibility to achieve an independent tuning of both QDs and cavity modes using Stark effect and electrostatic actuation within the same device and coupling to waveguides, towards the realization of efficient and scalable quantum photonic integrated circuits.

We acknowledge fruitful discussions with H. P. M. M. Ambrosius and R. W. van der Heijden. This research is supported by the Dutch Technology Foundation STW, applied science division of NWO, the Technology Program of the Ministry of Economic Affairs under project No. 10380, the FOM project No. 09PR2675 and the State of Bavaria.

\end{document}